\documentclass[15pt,a4paper]{article}
\usepackage{epsfig,amsmath,amssymb}
\usepackage[colorlinks=true,linktocpage=true,linkcolor=blue,citecolor=blue]{hyperref}
\usepackage{color}
\newcommand{\be}{\begin{equation}}
\newcommand{\ee}{\end{equation}}
\newcommand{\ba}{\begin{eqnarray}}
\newcommand{\ea}{\end{eqnarray}}

\newcommand{\nd}{\noindent}
\begin{document}
\begin{flushright}
{\normalsize
}
\end{flushright}
\vskip 0.1in
\begin{center}
{\large{\bf 
Heavy quark potential in a static and strong homogeneous magnetic field}}
\end{center}
\vskip 0.1in
\begin{center}
Mujeeb Hasan\footnote{hasan.dph2014@iitr.ac.in}, Bhaswar 
Chatterjee\footnote{bhaswar23@gmail.com}
and Binoy Krishna Patra\footnote{binoyfph@iitr.ac.in} 

\vskip 0.1in
{\small {\it Department of Physics, Indian Institute of 
Technology Roorkee, India, 247 667} }
\end{center}
\vskip 0.01in
\addtolength{\baselineskip}{0.4\baselineskip} 


\begin{abstract}
We have investigated the properties of quarkonia in a 
thermal QCD medium in the background of strong magnetic field. For 
that purpose, we employ the Schwinger 
proper-time quark propagator in the lowest Landau level to 
calculate the one-loop gluon self-energy, which in the 
sequel gives the the effective gluon propagator. 
As an artifact of strong magnetic field
approximation ($eB>>T^2$ and $eB>>m^2$), the Debye mass 
for massless flavors is found to depend only on the magnetic 
field which is the dominant scale in comparison to the scales 
prevalent in the thermal medium. However, 
for physical quark masses, it depends on both magnetic field and 
temperature in a low temperature and high magnetic field but the 
temperature dependence is very meagre and becomes independent of
temperature beyond a certain temperature and magnetic field.
With the above mentioned ingredients, the potential between 
heavy quark ($Q$) and anti-quark ($\bar Q$) is obtained in a hot QCD 
medium in the presence of strong magnetic field by correcting both 
short and long range components of the potential in real-time formalism. 
It is found that the long range part of the quarkonium potential is 
affected much more by magnetic field as compared to the short range part. 
This observation facilitates us to estimate the magnetic field beyond which 
the potential will be too weak to bind $Q\bar Q$ together. For example, 
the $J/\psi$ is dissociated at $eB \sim$ 10 $m_\pi^2$ and 
$\Upsilon$ is dissociated at $eB \sim$ 100 $m_\pi^2$ whereas its 
excited states, $\psi^\prime$ and $\Upsilon^\prime$ are dissociated at 
smaller magnetic field $eB= m_\pi^2$,  $13 m_\pi^2$, respectively.
\end{abstract}



\vspace{1mm}

\section{Introduction}
Lattice gauge theory at very high temperatures and/or baryon 
densities predicts an interesting window onto the properties
of Quantum Chromodynamics (QCD) in guise of a new 
phase, Quark-gluon Plasma (QGP), which pervaded the early universe, 
and may be present in the core of neutron stars. To realize this 
predicted phase, current 
experimental program of ultra relativistic heavy ion collisions 
(URHIC) have been designed at different colliders with different center 
of mass energies, {\em viz.} Relativistic Heavy Ion 
Collider (RHIC) at  Brookhaven National Laboratory 
(BNL) at $\sqrt{s}$= 200 GeV per 
nucleon in Au + Au collisions and Large Hadron 
Collider (LHC) at  European Organization for Nuclear 
Research (CERN) at $\sqrt{s}$= 2.76 TeV per nucleon 
in Pb + Pb collisions. Recent analysis suggests that 
the events of URHIC should be analyzed by 
incorporating the effect of magnetic field because an
intensely strong magnetic field, perpendicular to the 
reaction plane, is expected to be produced at very early 
stages of collisions when the event is off-central
\cite{Shovkovy:LNP871_2013,Elia:LNP871_2013,
Fukushima:LNP871_2013,Mulller:PRD89_2014,
Miransky:PR576_2015}. 
Depending on the centrality, the strength of the magnetic field may 
reach between $m_{\pi}^2$ ($\simeq 10^{18}$ Gauss) at 
RHIC \cite{Kharzeev:NPA803_2008} to 10 $m_{\pi}^2$ at 
LHC \cite{Skokov:IJMPA24_2009}. 
At extreme cases it may reach values of 50 $m_{\pi}^2$. 
A very strong magnetic field ($\sim 10^{23}$~Gauss) was also 
produced in the early universe during the electroweak
phase transition due to the gradients in Higgs field
\cite{vachaspati:PLB265'1991}. 
Ultimately such strong magnetic field might 
significantly affect the production of particles and 
alter their dynamics at very early stage of the collisions. 
Since magnetic field induces an anisotropy to the 
momentum of the affected particles, we might expect it 
to affect the anisotropic flow of the particles.

Naive classical estimates predict that the magnetic field may be 
very strong typically up to $t_{B} \simeq 0.2$ fm \cite{Mclerran:NPA929_2014}. 
However, the realistic calculations on the charge transport properties
of the plasma (namely, conductivity) may suggest that the magnetic 
field may remain substantial for significantly longer 
time~\cite{Tuchin:PRC82_2010}. Simultaneously heavy quark and antiquarks
pairs also develop into a physical resonances over a formation time
$t_{\rm form}\sim 1/E_{\rm bind}$ related to the binding energy of the 
state, {\em e.g.} the charm-anti charm ($ c \bar c$) pairs form resonances at
$t_{c \bar c} \sim 0.3 $ fm. Thus it becomes 
reasonable to assume that charmonium production may get 
significantly influenced by the strong magnetic field. The same 
argument applies to the bottomonium production. 
A large number of studies on the in-medium properties of 
$Q \bar Q$ bound states has been carried out using the 
phenomenological potential models \cite{Karsch:ZPC37_1988,Binoy:PLB505:2001,
Kakade:PRC92'2015},  
where the effects of the medium are encoded in a 
temperature dependent potential with non perturbative 
inputs from the lattice simulations. However, lattice calculations of 
free energies and other quantities \cite{McLerran:PRD24_1981} 
obtained from the correlation functions of Polyakov loop 
are often taken as input for the potential. Although these 
quantities have been thought to be related to the 
color-singlet and color-octet heavy quark potentials at 
finite temperature, a precise answer is still missing. 
Recently quarkonia 
at finite temperature has been studied by taking the advantage of 
the hierarchies between the non-relativistic scales 
associated with quarkonia and the thermal energy scales 
characterizing the system through the effective field 
theories, {\em viz. NRQCD, pNRQCD} 
etc. \cite{Brambilla:PRD78_2008}. The in-medium modifications 
of the quarkonium states can be studied from the
first principle of QCD by the spectral
functions \cite{Alberico:PRD77_2008} but the reconstruction of the
spectral function from the lattice meson correlators turns 
out to be very difficult. Recently in a new theoretical developments, the heavy
quark potential have also been synthesized in strong coupling regime
through a novel idea of gauge-gravity duality
~\cite{Maldacena:ATMP2'1998,Binoy:PRD92'2015,Binoy:PRD91'2015}.

Recently the properties of quarkonia states in a hot 
medium are explored in perturbative thermal QCD framework 
by correcting both the perturbative and non perturbative 
terms of the $Q \bar Q$ potential through the dielectric 
function in real-time formalism \cite{Thakur:PRD88_2013,Thakur:PRD89_2014} 
in both isotropic as well as anisotropic hot QCD medium, where
the anisotropy in the momentum space arose at the very early stages 
of the collisions due to the different expansion rate in the longitudinal
and transverse direction \cite{Kakade:IJMPA30_2015}. As mentioned earlier, 
magnetic field is also produced at the early stages of the collisions, thus 
it becomes worthwhile to examine the effects of magnetic field on the 
properties of quarkonia bound states, which is the central theme of our 
present work. Quantum mechanically both the quarkonium and heavy 
meson spectra have been analyzed through the solution
of non relativistic Schr\"{o}dinger equation with both harmonic oscillator 
and Cornell potential with an additional spin-spin 
interaction term \cite{Strickland:PRD88_2013,Bonati:PRD92_2015}.
Moreover lattice studies have also recently explored the possible anisotropies 
emerging in the static quark-anti quark 
potentials both at $T = 0$ and $T \ne 0$ through the
Wilson loop expectation value and Polyakov loop correlator, respectively,
in the presence of a magnetic 
background with respect to the direction of magnetic field
\cite{Bonati:PRD89_2014,Bonati:PRD94_2016}.

Here we have tried to investigate the effect of strong and homogeneous 
magnetic field on the properties of quarkonia states. We have first 
calculated the gluon self energy at finite temperature in a strong magnetic 
background and then obtain the heavy quark potential by taking the
static limit of the effective gluon propagator to see the effects
of magnetic field alone on the quarkonium states even in a thermal medium.
This is due to the fact that the magnetic field is assumed much stronger 
than the temperature as well as the mass of the quarks in quark-loop
of gluon self-energy ($eB \gg  T^2$ and $eB \gg m^2$), known as
``Strong Magnetic Field Approximation (SMFA)". As a consequence the Debye 
mass obtained from the static limit of gluon self-energy becomes almost 
independent of temperature, hence the potential even in the thermal
medium depends mainly on the magnetic field because the medium dependence
in the potential enters through the Debye mass. 
Moreover there is another condition for the non relativistic 
potential approach for heavy quarkonia to be valid is that the mass of 
the heavy quark (either charm or bottom quark) should be larger than the
dominant scale available in the problem ($m_Q>>\sqrt{eB}$)
(dimensionally) because $\sqrt{eB}$ is the most dominant scale
available in the strongly magnetized thermal medium. Thus the 
above mentioned two conditions constraint the lower and upper limit 
of the magnetic fields, respectively, for which our work is valid. 
As a bi-product of this constraint, the magnetic field expected to be 
produced at RHIC may not satisfy the condition of SMFA. So our work
which is valid only in SMFA will be suitable to the LHC
events where the magnetic field expected to be 
produced are well within the limit of the validity
of our work.

Our work is arranged in the following way: 
In subsection 2.1 we will discuss the quark propagator at finite
temperature within SMFA. In subsection 2.2 
we will calculate the gluon self energy at finite temperature
in presence of strong magnetic field. In subsection 2.3 we will compute 
the screening mass in SMFA by taking the static limit of gluon-self energy.
In Section 3, we will obtain the potential from the inverse Fourier transform
of the effective propagator in the static limit 
and explore how the properties of quarkonia
could be affected by the presence of strong magnetic field.
Finally we will conclude in Section 4.

\section{One-loop gluon self-energy and the screening mass in SMFA}

 The gluon self energy can be affected by the magnetic field 
in two ways: First, the quark propagator gets affected due to 
the Landau quantization of the energy levels (known as Landau levels) 
in the presence of magnetic field. Second, the strong coupling runs 
with both the magnetic field and temperature. However, in SMFA, it 
runs exclusively with the magnetic field as we have discussed 
in the introduction.

Schwinger first obtained the fermionic propagator in coordinate space 
by the proper-time formalism \cite{Schwinger:PR82_1951}, then Tsai has 
obtained the same in momentum space and used it to calculate the 
vacuum polarization in magnetic field \cite{Tsai:PRD10_1974}. The vacuum
polarization tensor has also been obtained in a gauge invariant
manner for both strong and weak magnetic field limit 
\cite{Shabad:AP121_1979,Palash:PRD60_1999,Sarira:PRD67_2003}.
We shall now going to extend these calculations to QCD to calculate
the gluon self-energy, which will in turn help to study
the properties of quarkonia quantum field theoretically in
the presence of strong magnetic field.

\subsection{Fermionic propagator in presence of magnetic field}
\subsubsection{ Vacuum in a static and homogeneous magnetic 
field:} 
For the sake of simplicity, we assume the magnetic field 
to be constant and homogeneous. We also assume the magnetic 
field to be along $z$-direction and of magnitude 
$B$. Such a magnetic field can be 
obtained from a vector potential $A_\mu = (0,0,Bx,0)$. 
The choice of vector potential is not unique as the same magnetic 
field can also be obtained from a symmetric potential given by 
$A_\mu = (0,\frac{-By}{2},\frac{Bx}{2},0)$.
Using the proper-time method formulated by Schwinger, the fermion 
propagator in such a magnetic field can be written in the coordinate 
space as \cite{Schwinger:PR82_1951}
\begin{equation}
S(x,x^\prime) = \phi (x,x^\prime)\int \frac{d^4 p}{(2\pi)
^4}e^{ip(x-x^\prime)}S(p)~,
\label{schwinger}
\end{equation}
where the the phase factor, $\phi (x,x^\prime)$ is given by
\begin{equation}
\phi (x,x^\prime) = e^{ie\int_x^{x^\prime}A(\xi)d\xi}~,
\label{phase}
\end{equation}
which becomes unity for a closed fermion loop with two fermion lines, 
i.e, $\phi(x,x^\prime)=1$ \cite{Chyi:PRD62_2000}.\\
However, the same propagator was first calculated by \cite{Tsai:PRD10_1974,
Chyi:PRD62_2000} in 
the momentum space as 
\begin{equation}
iS(p)=\int_0^\infty \frac{1}{eB}\frac{ds}{\cos(s)}
e^{-is\left[m^2-p_{\|}^2+\frac{\tan(s)}{s}
p_\bot^2\right]}
\left[\left\lbrace\cos(s) +\gamma_{1}\gamma_{2} 
\sin(s)\right\rbrace(m+\gamma\cdot p_{\|})
-\frac{\gamma\cdot p_{\bot}}{\cos (s)}\right],
\label{sch_mom}
\end{equation}

The propagator (\ref{sch_mom}) in the momentum space can also be 
expressed in a more convenient way using the associated Laguerre 
polynomials ($L_n$)
\begin{equation}
iS(p)=\sum_n\frac{-id_n(\alpha)D+d^\prime_n(\alpha)
\bar{D}}{p_L^2+2neB}
+i\frac{\gamma\cdot p_\bot}{p_\bot^2}~,
\label{prop_lag}
\end{equation}
where the following quantities are defined as \cite{Chyi:PRD62_2000}
\begin{eqnarray*}
D&=&(m+\gamma\cdot p_\|)+\gamma\cdot 
p_\bot\frac{m^2-p_\|^2}{p_\|^2},\\
\bar{D}&=&\gamma_{1}\gamma_{2}(m+\gamma\cdot p_\|),\\
d_n(\alpha)&=&(-1)^n e^{-\alpha}C_n(2\alpha),\\ 
C_n(2\alpha)&=&L_{n}(2\alpha)-L_{n-1}(2\alpha),\\
d^{'}_{n}(\alpha)&=&\frac{\partial d_n}{\partial\alpha},\\ 
p_L^2&=&m^2-p_\|^2,\\
\alpha &=&\frac{{p_\bot}^2}{eB},\\
p_{\|}^2&=&p_{0}^2-p_{z}^2, \\
p_{\bot}^2&=&p_{x}^2+p_{y}^2~.
\end{eqnarray*}
The order of the Laguerre polynomial also 
corresponds to the number of energy eigenvalues in a magnetic field,
known as Landau levels. In SMFA, the particles occupy 
the lowest Landau level (LLL) ($n=0$) only, thus, in SMFA, the 
fermion propagator in eq.(\ref{prop_lag}) reduces to the following form 
\begin{equation}
iS_0(p)=\frac{(1+\gamma^{0}\gamma^{3}\gamma^{5})
(\gamma^{0}p_{0}-\gamma^{3}p_{z}+m)}
{p_{\parallel}^2-m^2+i\epsilon} e^{-\frac{p_{\perp}^2}
{\mid qB\mid}},
\label{vacrop}
\end{equation}
where $m$ and $q$ are the mass and electric charge  of the fermion, 
respectively.

\subsubsection{Heat Bath in a strong homogeneous magnetic field}
In thermal medium, the system possesses additional thermal scales, 
{\em viz.} T, $gT$ etc., which are well separated in weak coupling regime 
($T>gT>\cdot \cdot$), in addition to the quark masses. So at finite 
temperature, strong magnetic field approximation implies 
that both conditions $eB>>T^2$ and $eB>>m^2$ are to be satisfied. 
To switch on the temperature in the vacuum propagator 
(\ref{vacrop}) in real-time formalism, the matrix propagator is 
diagonalized by the matric $U$ as
\begin{equation}
S_{ab}(p)=U_{ac} (p)
{\begin{pmatrix} 
  S_0(p)     & 0\\ 
  0 &-S_0^*(p)  
\end{pmatrix}}_{cd} U_{db}(p),
\label{propagator_matrix}
\end{equation}
where the matrix U(p) is given by
\begin{equation}
U(p)=
\begin{pmatrix} 
 \tilde N_2     & -\tilde N_1 e^{-\beta \mu/2}\\ 
  \tilde N_1 e^{-\beta \mu/2}&  \tilde N_2
\end{pmatrix},
\label{u_matrix}
\end{equation}
with
\begin{eqnarray}
\tilde{N_1}(p_0)&=&\sqrt{n^{+}_p}~\theta(p_0)+ 
\sqrt{n^{-}_p}~\theta(-p_0),\\
\tilde{N_2}(p_0)&=&\sqrt{1-n^{+}_p}~\theta(p_0)+
\sqrt{1-n^{-}_{p}}~\theta(-p_0),\\
n^{\pm}_p (p_0)&=&\frac{1}{e^{\beta(p_0\mp\mu)+1}},
\end{eqnarray}
where $\beta$ is the inverse of temperature and $\mu$ is the 
chemical potential. In the present work we are working for baryonless
medium ($\mu=0$), {\em i.e.}  $n^{+}_p = n^{-}_p = n_p$. 
Plugging eq.(\ref{u_matrix}) in eq.(\ref{propagator_matrix}), 
we get the fermion propagator as 
\begin{equation}
S(p) =
\begin{pmatrix}
S_0(p)\tilde N_2^{2}+S_0^{*}(p)\tilde N_1^{2}& 
-S_0(p)\tilde{N_1}\tilde{N_2}+S_0^{*}(p)\tilde{N_1}\tilde{N_2} \\
S_0(p)\tilde{N_1}\tilde{N_2}-S_0^{*}(p)\tilde{N_1}\tilde{N_2}& -
S_0(p)\tilde N_1^{2}-S_0^{*}(p)\tilde N_2^{2} 
\end{pmatrix}~.
\label{temp_prop}
\end{equation}

For calculating the gluon self energy in an equilibrium medium,
we need only the ``11''-component of the matrix propagator expressed 
in eq.(\ref{temp_prop})
\begin{equation}
S_{11}(p)=S_0(p)\tilde N_2^2+S_0^{*}(p)
\tilde N_1^2 ~,
\end{equation}
\begin{equation}
iS_{11}(p)=\Bigg[\frac{1}{{p_{\parallel}^2-m^2+
i\epsilon}}+2\pi in_{p}\delta(p_{\parallel}^2-m^2)\Bigg]
(1+\gamma^{0}\gamma^{3}\gamma^{5})(\gamma^{0}p_{0}-\gamma^{3}
p_{z}+m) e^{\frac{-p_{\perp}^2}{\mid qB \mid}},
\end{equation}
where the distribution function is given by 
\begin{equation*}
n_p(p_0) = \frac{1}{e^{\beta\mid p_0\mid} + 1}.
\end{equation*}
The above description for fermionic propagator can be easily generalized to 
quarks of $f$-th flavor with which we are going to calculate the gluon 
self-energy. 

\subsection{Gluon self-energy in a hot QCD medium in presence of strong 
magnetic field}
 As mentioned earlier, since we are working within SMFA so
we may consider the strong coupling to run with the magnetic field  only.
For this 
purpose we closely follow the running coupling in \cite{Ferrer:PRD91_2015}, 
where the coupling runs with the momentum parallel and perpendicular to the 
magnetic field separately. In our case of magnetic field ($\vec{B}= B\hat{z}$), 
we will use the coupling dependent on the longitudinal component only because 
the energy of Landau levels for quarks in SMFA depend only on the longitudinal 
component of momentum. In fact, the coupling dependent on the transverse 
momentum does not depend on magnetic field at all. So in our calculation, the 
relevant coupling is given by~\cite{Ferrer:PRD91_2015}

\begin{equation}
\alpha_{s}^\|(k_3)=\frac{1}{{\alpha_s^0(\mu_0)}^{-1}+\frac{11N_c}{12\pi}
\ln(\frac{k^2_3+M^2_B}{\mu_0^2})+\frac{1}{3\pi}\sum_f \frac{|q_f B|}{\sigma}}, 
\end{equation}
where 
\begin{equation}
\alpha_s^0(\mu_0) = \frac{12\pi}
{11N_c\ln(\frac{(\mu_0^2+M^2_B)}{\Lambda_V^2})}.  
\end{equation}
In the above eq.(13), $M_B$ is taken $\sim~1$ GeV as an infrared mass and 
the string tension, $\sigma=0.18 {\rm{GeV}}^2$.

For system in equilibrium, we need only the ``11''-component 
of the gluon self-energy matrix calculated in real time formalism, which is given by
\begin{eqnarray}
{\Pi^{\mu\nu}}(k)&=& \frac{ig^2}{2} \sum_f \int\frac{d^{4}p}
{(2\pi)^4}tr[\gamma^{\mu}S_{11}(p)\gamma^{\nu}S_{11}(q)]\nonumber \\ 
&=&\frac{ig^2}{2} \sum_f \int\frac{d^{4}p}{(2\pi)^4}tr
[(\gamma^{\mu}(1+\gamma^{0}\gamma^{3}\gamma^{5})
(\gamma^{0}p_{0}-\gamma^{3}p_{z}+m_f)
\gamma^{\nu}(1+\gamma^{0}\gamma^{3}\gamma^{5})
(\gamma^{0}q_{0}-\gamma^{3}q_{z}+m_f)]\nonumber \\
&&\left\lbrace \frac{1}{p_{\parallel}^2-m_f^2+i\epsilon}
+2\pi i n_{p}\delta(p_{\parallel}^2-m_f^2)\right\rbrace \nonumber \\
&&\left\lbrace \frac{1}{q_{\parallel}^2-m_f^2+i\epsilon}
+2\pi i n_{q}\delta(q_{\parallel}^2-m_f^2)\right\rbrace
e^{\frac{{-p_{\perp}}^2}{|q_f|B}}e^{\frac{{-q_{\perp}}^2}{|q_f|B}},
\label{pmn}
\end{eqnarray}
where the factor 1/2 arises due to trace in color-space and the trace due to
$\gamma$ matrices is given by
\begin{equation}
L^{\mu\nu}=8\left[p_{\parallel}^{\mu}q_{\parallel}
^{\nu}+p_{\parallel}^{\nu}q_{\parallel}^{\mu}
-g_{\parallel}^{\mu\nu}((p.q)_{\parallel}-m_f^2)
\right].
\label{trace}
\end{equation}
Separating the momentum integration into longitudinal ($\parallel$) 
and transverse ($\perp$) components with respect to the magnetic
field, the gluon self-energy 
can be factorized into $\parallel$ and $\perp$ components of momentum 
integration
\begin{equation}
\Pi^{\mu\nu}(k)=\sum_f \Pi_{\parallel}^{\mu\nu}(k_{\parallel})A_f(k_{\perp})~,
\label{selfenergy}
\end{equation}
where the transverse component is given by
\begin{eqnarray}
A_f(k_{\perp})&=&\int dp_{x} dp_{y} e^{\frac{{-p_{\perp}}^2}
{|q_{f}|B}}e^{\frac{{-q_{\perp}}^2}{|q_{f}|B}}\nonumber\\
&=&\frac{\pi |q_{f}|B}{2}e^{-\frac{k_{\perp}^2}{2|q_{f}|B}}.
\label{pitr}
\end{eqnarray}
It may be noted that in LLL approximation, the dependence of self-energy 
on the magnetic field is fully encapsulated in the transverse component
whereas the longitudinal part carries no dependence on the magnetic field.
We will now calculate  the longitudinal component of the self-energy
by decomposing eq.(\ref{pmn}) into vacuum and thermal parts:
\begin{equation}
\Pi^{\mu\nu}_{\parallel}=(\Pi^{\mu\nu}_{\parallel})_{V}+
(\Pi^{\mu\nu}_{\parallel})_{n}+(\Pi^{\mu\nu}_{\parallel})_{n^2},
\end{equation}
where  $(\Pi^{\mu\nu}_{\parallel})_{V}$ is the vacuum 
part, $(\Pi^{\mu\nu}_{\parallel})_{n}$ and 
$(\Pi^{\mu\nu}_{\parallel})_{n^2}$ are the thermal 
contributions due to single and double distribution functions,
respectively. They are explicitly given by
\begin{eqnarray}
(\Pi^{\mu\nu}_{\parallel})_{V}&=&\frac{ig^2}{2(2\pi)^4}
\int dp_0 dp_z L^{\mu\nu}\left\lbrace
\frac{1}{(q_{\parallel}^2-m_f^2+i\epsilon)}\frac{1}
{(p_{\parallel}^2-m_f^2+i\epsilon)}\right\rbrace,
\label{pmnvac}\\
(\Pi^{\mu\nu}_{\parallel})_{n}&=&\frac{ig^2(2\pi i)}
{2(2\pi)^4}\int dp_0 dp_z L^{\mu\nu}\left\lbrace 
\frac{n_{p}\delta(p_{\parallel}^2-m_f^2)}{(q_{\parallel}^2
-m_f^2+i\epsilon)}
+\frac{n_{q}\delta(q_{\parallel}^2-m_f^2)}
{(p_{\parallel}^2-m_f^2+i\epsilon)}\right
\rbrace, 
\label{Pin1}\\
(\Pi^{\mu\nu}_{\parallel})_{n^2}&=&\frac{ig^2}{2(2\pi)^4}
\int dp_0 dp_z L^{\mu\nu}\lbrace 
(-4\pi^2)n_{p}n_{q}\delta(p_{\parallel}^2-m_f^2)
\delta(q_{\parallel}^2-m_f^2)\rbrace~. 
\label{Pin21}
\end{eqnarray}

We will now calculate the vacuum term for the gluon self-energy.

\subsubsection{Vacuum contribution ($T=0$, $eB \neq 0$)}

The vacuum term in strong magnetic field can be calculated easily 
as it is similar to the calculation of self energy in vacuum without 
magnetic field 
except the fact that the dimension of the momentum integration is 
now reduced  
from $4$ to $2$. This dimensional reduction in fact removes 
the divergences usually encountered in 4-dimension, thus
we do not need any regularization any more. Using the identity
\begin{equation}
\frac{1}{x \mp i\epsilon}=\textit{\textbf{P}}
\left(\frac{1}{x}\right)\pm i \pi \delta(x), 
\end{equation}
the real part of the vacuum term in the gluon-self energy has
been calculated as 
\begin{equation}
\Re {\Pi^{\mu\nu}(k) \mid}_V=\Big(g_{\parallel}^{\mu\nu}
-\frac{k_{\parallel}^{\mu}k_{\parallel}^{\nu}}{k_{\parallel}^2}
\Big)\Pi(k^2),
\end{equation}
where the form factor, $\Pi (k^2)$ is given by 
\begin{eqnarray}
\Pi(k^2)=\frac{g^2}{4\pi^2}\sum_{f}\mid q_{f}B \mid 
e^{-\frac{k_{\perp}^2}{2|q_{f}|B}}\left[\frac{2m_{f}^2}
{k_{\parallel}^2}
\left(1-\frac{4m_{f}^2}{k_{\parallel}^2}\right)^{-1/2}
\left\lbrace \ln\frac{1-{\Big(1-\frac{4m_{f}^2}
{k_{\parallel}^2}\Big)}^{1/2}}
{1+{\Big(1+\frac{4m_{f}^2}{k_{\parallel}^2}
\Big)}^{1/2}} + i\pi\right\rbrace -1\right].
\end{eqnarray}
Therefore the ``00''-component ($\mu=\nu=0$) of the real part of 
vacuum term of the gluon self-energy (using the metric
$g^{\mu\nu}_{\parallel}=diag(1,0,0,-1)$ is given by 
\begin{equation*}
\Re {\Pi^{00}(k)\mid}_{V}=-\frac{k_{z}^2}{k_{\parallel}^2}
~\Pi(k^2).
\end{equation*}
In the limit of massless quarks $(m_{f}=0)$, 
the gluon self-energy due to vacuum term in the 
static limit ($k_0=0$, $\vec{k} \rightarrow 0$) 
is given by the scale available to the magnetic field only in SMFA 
\begin{equation}
\Re {\Pi^{00} (k_0=0,\vec{k}\rightarrow 0)\mid}_V
=\frac{g^2}{4\pi^2}\sum_{f}\mid q_{f}B \mid.
\label{Pi00masslessb}
\end{equation}
For the physical quark masses $(m_{f}\neq0$), the vacuum term
in the static limit ($k_0=0$, $\vec{k} \rightarrow 0$) vanishes
\begin{equation}
\Re {\Pi^{00} (k_0=0,\vec{k}\rightarrow 0)\mid}_V=0.
\label{Pi00massive}
\end{equation}

\subsubsection{Medium contribution}
The (thermal) medium contribution to the gluon self-energy contains two 
terms: the first one (\ref{Pin1}) involves single distribution function 
and the second one (\ref{Pin21}) involves the product of two distribution 
functions. We will first consider the medium contribution due to
the single distribution function only. Using the property of Dirac delta 
function, the gluon self-energy in eq.(\ref{Pin1}) is reduced to
	\begin{eqnarray}
	(\Pi^{\mu\nu}_{\parallel})_{n}&=&-\frac{g^2}{2(2\pi)^3}
	\int dp_0 dp_z L^{\mu\nu}\Bigg[
	\frac{n_{p}(p_0)\Big\{\delta(p_{0}-\omega_{p})
	+\delta(p_{0}+\omega_{p})\Big\}}{(q_{0}^2-q_{z}^2-m_f^2
	+i\epsilon)(2\omega_{p})}\nonumber \\
	&+&\frac{n_{q}(q_0)\Big\{\delta(q_{0}-\omega_{q})
	+\delta(q_{0}+\omega_{q})\Big\}}{(p_{0}^2-p_{z}^2-m_f^2
	+i\epsilon)(2\omega_{q})}\Bigg].
	\end{eqnarray}
Taking $\mu=\nu=0$, the real part of ``00''-component of 
$(\Pi^{\mu \nu}_{\parallel})_{n}$ becomes 
	\begin{eqnarray}
	\Re {\Pi^{00}_\parallel (k_0,k_z)\mid}_n &=&-\frac{g^2}
	{2(2\pi)^3}\int dp_0 dp_z L^{00}\Bigg[
	\frac{n_{p}(p_0)\Big\{\delta(p_{0}-\omega_{p})
	+\delta(p_{0}+\omega_{p})\Big\}}{(q_{0}^2-\omega_q^2)
	(2\omega_{p})}\nonumber \\
	&+&\frac{n_{q}(q_0)\Big\{\delta(q_{0}-\omega_{q})
	+\delta(q_{0}+\omega_{q})\Big\}}{(p_{0}^2-\omega_p^2)
	(2\omega_{q})}\Bigg],
	\label{pn00}
	\end{eqnarray}
where the ``00'' component of $L^{\mu\nu}$ is
	\begin{equation}
	L^{00}=8[p_{0}q_{0}+p_{z}q_{z}+m_f^{2}],
	\label{trace}
	\end{equation}
and the other notations are 
	\begin{eqnarray*}
	\omega_p&=&\sqrt{p_z^2+m_f^2},\\
	\omega_q&=&\sqrt{(p_z-k_z)^2+m_f^2}.
	\end{eqnarray*}
After performing the $p_{0}$ integration we get from  eq.(\ref{pn00})
	\begin{eqnarray}
\Re {\Pi^{00}_\parallel (k_0,k_z)\mid}_n&=&-\frac{g^2}{4(2\pi)^3}\int dp_z\Bigg
	[\frac{L^{00}_{1}~n_p^{+}}{\omega_p[(\omega_p-k_0)^2-\omega_q^2]}
	+\frac{L^{00}_{2}~n_p^{-}}{\omega_p[(\omega_p+k_0)^2-\omega_q^2]}\nonumber \\
	&+&\frac{L^{00}_{3}~n_q^{+}}{\omega_q[(\omega_q+k_0)^2-\omega_p^2]}
	+\frac{L^{00}_{4}~n_q^{-}}{\omega_q[(\omega_q-k_0)^2-\omega_p^2]}
	\Bigg] \label{pin},
	\end{eqnarray}
	where we have defined
	\begin{eqnarray*}
	L^{00}_{1}&=&L^{00}(p_0=\omega_p)=
	8(2\omega_p^2-\omega_{p}k_0-p_{z}k_{z}),\\
	L^{00}_{2}&=&L^{00}(p_0=-\omega_p)=
	8(2\omega_p^2+\omega_{p}k_0-p_{z}k_{z}),\\
	L^{00}_{3}&=&L^{00}(p_0=\omega_q+k_0)=
	8(2\omega_p^2+\omega_{q}k_0-3p_{z}k_{z}+k_{z}^2),\\
	L^{00}_{4}&=&L^{00}(p_0=-\omega_q+k_0)=
	8(2\omega_p^2-\omega_{q}k_0-3p_{z}k_{z}+k_{z}^2),
	\end{eqnarray*}
	and 
	\begin{eqnarray*}
	n_p^{+}&=&n_p(p_0=\omega_p),\\
	n_p^{-}&=&n_p(p_0=-\omega_p),\\
	n_q^{+}&=&n_q(p_0=\omega_q+k_0),\\
	n_q^{-}&=&n_q(p_0=-\omega_q+k_0).
	\end{eqnarray*}

In the limit of massless quarks ($m_{f}=0$), the gluon self-energy 
in eq.(\ref{pin}) gets simplified into
	\begin{equation}
	\Re {\Pi^{00}_\parallel (k_0,k_z)\mid}_n=\frac{8g^2}{2(2\pi)^3} 
	\Bigg[\frac{k_z^2}{k_0^2-k_z^2}+\frac{k_{z}T}{k_0^2-k_z^2}\ln(2)
	-\frac{k_{z}T}{k_0^2-k_z^2}\ln(1+e^{\frac{k_z}{T}})\Bigg].
	\end{equation}
Using eq.(\ref{selfenergy}) and  multiplying the transverse component,
$A(k_\perp)$ from eq.(\ref{pitr}), the contribution to the real
part of self-energy from the component having single distribution function 
becomes
	\begin{equation}
	\Re {\Pi^{00} (k_0,k_x,k_y,k_z)\mid}_n=\frac{g^2}{4\pi^2}\sum_{f} 
	|q_{f}|B e^{-\frac{(k_{x}^2+k_{y}^2)}{2|q_f|B}}
	\Bigg[\frac{k_z^2}{k_0^2-k_z^2}+\frac{k_{z}T}{k_0^2-k_z^2}\ln(2)
	-\frac{k_{z}T}{k_0^2-k_z^2}\ln(1+e^{\frac{k_z}{T}})\Bigg]~,
	\end{equation}
which, in the static limit ($k_0=0$, $\vec{k} \rightarrow 0$) becomes
	\begin{equation}
	\Re {\Pi^{00}(k_0=0,\vec{k}\rightarrow 0) \mid}_n
	=-\frac{g^2}{4\pi^2}\sum_{f} 
	|q_{f}|B+\frac{g^2}{8\pi^2}\sum_{f} 
	|q_{f}|B.
	\label{d}
	\end{equation}

\nd However, for the physical quark masses $(m_{f}\neq0$),
the self-energy in eq.(\ref{pn00}) reduces to, by putting $k_0=0$
	\begin{eqnarray}
\Re {\Pi^{00}_\parallel (k_0=0,k_z) \mid}_n=-\frac{g^2}{2(2\pi)^3}\int  dp_z 
I_{n},
	\end{eqnarray}
	where the integrand, $I_n$, is given by
	\begin{equation}
	I_{n}=\frac{8p_{z}n_p}{\omega_{p}k_{z}}
	-\frac{8(p_{z}-k_{z})n_q}{\omega_{q}k_{z}}+\frac{16m_{f}^2n_p}
	{\omega_{p}k_{z}(2p_{z}-k_{z})}-\frac{16m_{f}^2n_q}
	{\omega_{q}k_{z}(2p_{z}-k_{z})},
	\end{equation}
and the distribution functions are given by
	\begin{equation*}
	n_{p}=\frac{1}{e^{\beta|\omega_{p}|}+1}, \quad \quad n_{q}=\frac{1}
	{e^{\beta|\omega_{q}|}+1}.
	\end{equation*}
Further taking the $k_z \rightarrow 0$ limit, the integrand, $I_{n}$ is 
simplified into
	\begin{equation*}
	I_{n}=-\frac{8}{T}n_{p}(1-n_{p}).
	\end{equation*}
Thus for the physical quark masses $(m_{f}\neq0$),
the contribution to the gluon self-energy having single
distribution function in the static limit reduces to
        \begin{equation}
	\Re {\Pi^{00} (k_0=0,\vec{k}\rightarrow 0) \mid}_n=
	\frac{g^2}{4\pi^{2}T}\sum_{f}|q_{f}B|\int_{0}^{\infty}dp_{z}
	\frac{e^{\beta\omega_{p}}}{(1+e^{\beta\omega_{p}})^2}.
	\label{f}
	\end{equation}

Finally the medium contribution to the gluon self-energy involving the product of 
two distribution functions given in eq.(\ref{Pin21})  
does not contribute to the real-part of the gluon self-energy, i.e. 
	\begin{equation}
	\Re {\Pi^{00} (k_0=0,\vec{k}\rightarrow 0)\mid}_{n^2}=0.
	\label{e}
	\end{equation}
We have thus so far evaluated the vacuum as well as medium contribution to
one-loop gluon self energy, therefore we add them up to obtain the 
real-part of one-loop gluon self-energy in static limit for massless quarks
	\begin{equation}
	\Re \Pi^{00}(k_0=0,\vec{k}\rightarrow 0)=
	\frac{g^2}{8\pi^2}\sum_{f} 
	|q_{f}|B ,
	\label{h}
	\end{equation}
and for the physical quark masses ($m_f \ne 0$)
	\begin{equation}
	\Re \Pi^{00}(k_0=0,\vec{k}\rightarrow 0)=\frac{g^2}{4\pi^{2}T}
	\sum_{f}|q_{f}B|\int_{0}^{\infty}
	dp_{z}\frac{e^{\beta\omega_{p}}}{(1+e^{\beta\omega_{p}})^2}.
	\label{g}
	\end{equation}

\subsection{Debye screening mass in strong magnetic field:}

The Debye screening manifests in the collective oscillation of 
the medium via the dispersion relation and is obtained by the 
static limit of the longitudinal part (``00'' component) 
of gluon self-energy, {\em i.e.}
\begin{equation}
m_{D}^2=\Re \Pi^{00} (k_0=0,\vec{k}\rightarrow 0). 
\end{equation}
Therefore, eq.(\ref{h}) gives the very simple form 
for the square of the Debye mass for massless quarks, which is already derived 
in \cite{Fuku:PRD93_2016,Munshi:PRD:94_2016}
\begin{equation}
m_{D}^2=\frac{g^2}{8\pi^2}\sum_{f}|q_{f}|B.
\end{equation}
It shows that $m_D^2$ depends strongly on the magnetic field 
and is independent of temperature, thus the collective behavior of 
the medium gets strongly affected by the presence of 
strong magnetic field. However, for physical quark masses, the 
Debye mass is given by from eq.(\ref{g})
\begin{equation}
m_{D}^2=m_{D}^2(m_f=0) \times \frac{2}{T}\int_{0}^{\infty}
dp_{z}\frac{e^{\beta\omega_{p}}}{(1+e^{\beta\omega_{p}})^2},
\label{md}
\end{equation}
which depends on both magnetic field and temperature. However, $m_D^2$
depends strongly on the magnetic field and the dependence on temperature is 
very weak and the screening mass becomes temperature-independent beyond a 
certain temperature.

Now, for the SMFA to be valid, we have to be careful in choosing 
the range of temperature and magnetic field. {\em For example}, for 
temperatures up to 300 MeV, the starting value of $eB$ has to 
be much higher than $0.09~{\rm{GeV}}^2$. Here we have taken the 
starting magnetic field to be $eB=10~m_\pi^2\sim 0.2~GeV^2$.
However, for the upper 
bound on the magnetic field, the constraint comes from the heavy quark 
mass ($m_Q \gg
\sqrt{eB}$) as discussed in the introduction. So, we have taken the 
highest magnetic field for charmonium states to be $eB=25~m_\pi^2$ which 
gives us $\sqrt{eB}\sim 0.7~GeV$. Thus, to see the variation of the Debye 
masses with the strong magnetic field,
we have numerically calculated $m_D^2$ as a function of $eB$ 
(in units of $m_\pi^2$) for the temperature range T=200 - 300 MeV 
in Fig.1-a and noticed that $m_D^2$ is almost linearly increasing with $eB$ for smaller 
temperature. For higher temperatures, $m_D^2$ deviates slightly from
the linearly increasing trend. 

\begin{figure}[t]
\begin{center}
\begin{tabular}{c c}
\includegraphics[width=5.5cm,height=5cm]{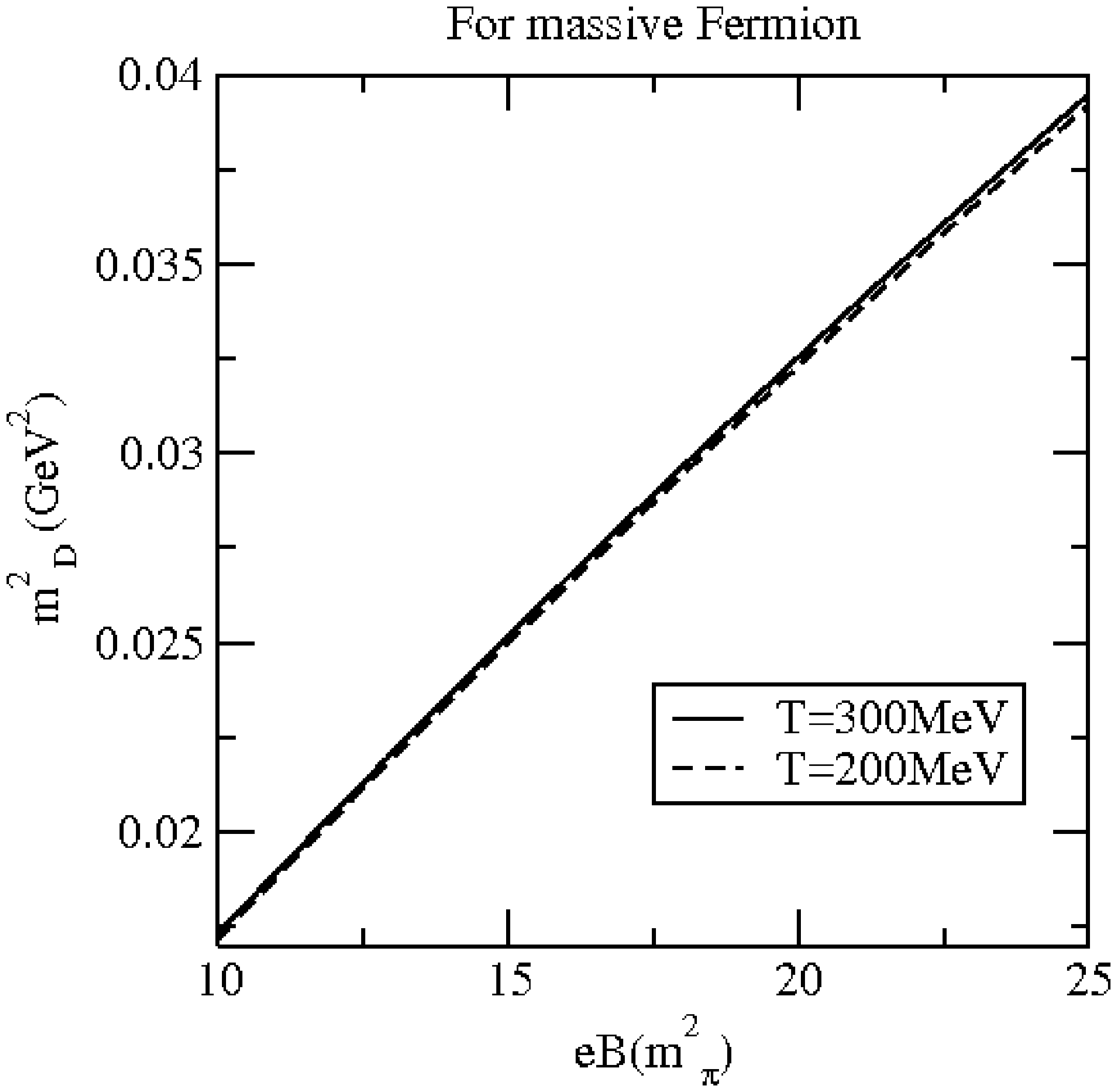}&
\includegraphics[width=5.5cm,height=5cm]{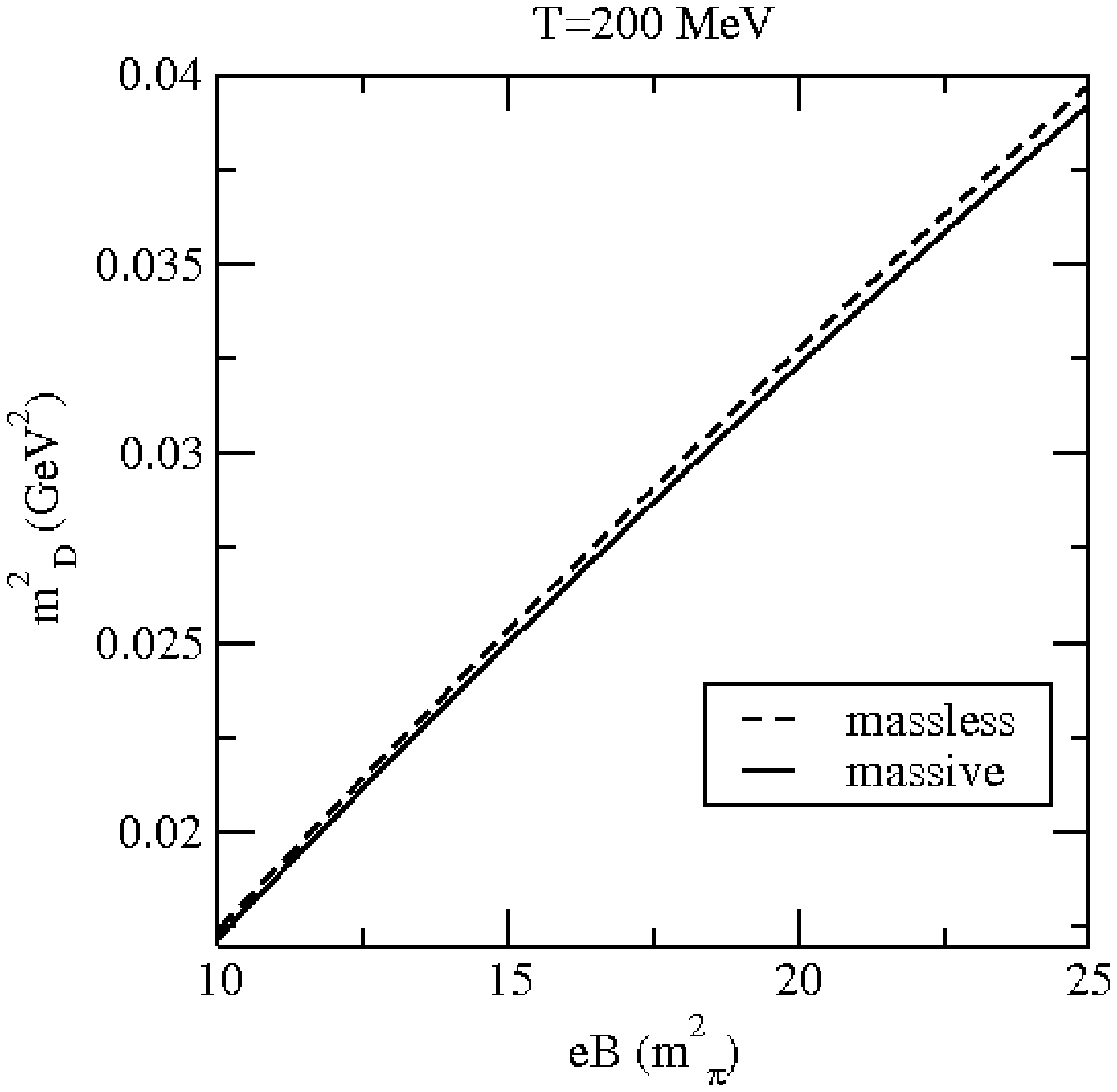}\\
a & b
\end{tabular}
\caption{\textbf{Left panel:}Separation is seen only between low and high T
at high eB. \textbf{Right panel:} High eB can distinguish b/w massless
and massive fermions (quarks).}
\end{center}
\label{fig1}
\end{figure}

As we understood earlier in SMFA, the strongly magnetized thermal medium 
with massless quarks possesses only one scale available related to the
magnetic field ($eB$) so by the dimensional arguments the square of the Debye
mass is linear in $eB$  whereas for the the medium with physical quark masses,
even in SMFA there is a weak competition between the dominant scale, $eB$ and
much weaker 
scales, mass (m) and temperature (T) (rather their ratio, $m/T$) in the 
form of Boltzmann damping factor ($\exp (-m/T)$) as in eq.(\ref{md}). This is seen 
in Fig.1-b, 
where a comparison of Debye masses with and without incorporating the 
quarks masses is made.

\begin{figure}[h]
\begin{center}
\includegraphics[width=7cm,height=5.9cm]{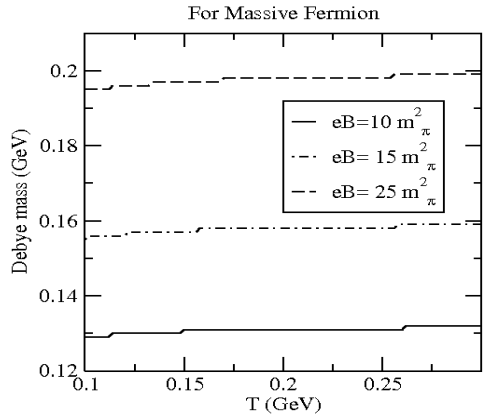}
\caption{Effect of temperature is only pronounced at low 
temperature and
high magnetic field.}
\end{center}
\label{fig2}
\end{figure}

To see the temperature dependence of the Debye mass 
explicitly we have plotted $m_D$ with the temperature directly with 
increasing values of $eB= 10 m_\pi^2$, $15 m_\pi^2$ and $25 m_\pi^2$ 
in Fig.2. We can see how weakly the screening mass depends 
on temperature. It increases very slightly with temperature and beyond a 
point, the screening mass is practically a constant with magnetic field. 
The effect of temperature is slightly more pronounced for 
high magnetic field and low temperature.

Very recently the effects of a magnetic background on color-screening 
phenomena in QGP is also explored through the estimation of both the 
magnetic and electric screening masses by measuring the Polyakov loop 
correlators on the lattice for various temperatures~\cite{Bonati:PRD95_2017}. 
There, it is found that the magnetic field induces an
increase of both the magnetic and the electric screening
masses and, to some extent, also the appearance of an
anisotropy in Polyakov loop correlators. Both screening masses
are found to increase linearly with the magnetic field and the influence
of the magnetic field on both the masses is enhanced at lower 
temperatures and is asymptotically diminished in the higher temperature.
Thus our aforesaid results on the Debye mass qualitatively agree with
their findings for the electric screening mass,
which is of interest to us for the screening of heavy quark potential.
However, their lattice estimates for the electric screening masses
are approximately larger by an order of magnitude than our results.
This large difference may be attributed due to the nonperturbative 
effects, which is beyond the scope of this work.   

\section{Heavy quark potential in a hot QCD medium:}
The derivation of potential between a heavy quark $Q$ and its anti-quark
($\bar Q$) either from EFT (pNRQCD) or from first
principle QCD may not be plausible because the hierarchy of non relativistic 
scales and thermal scales assumed in weak coupling EFT calculations may not 
be satisfied and the 
adequate quality of the data is not available in the present lattice
correlator studies, respectively, so one may use the potential model to 
circumvent the problem.

Since the mass of the heavy quark ($m_Q$) is very large, so the 
requirement: $m_Q \gg \sqrt{eB} \gg \Lambda_{QCD}$ and $T \ll m_Q$
is satisfied for the description of the interactions between a pair of heavy 
quark and anti-quark at finite temperature in the presence of magnetic 
field in terms of quantum mechanical 
potential. Thus we can obtain the medium-modification to the vacuum potential
by correcting both its short and long-distance part with a dielectric 
function $\epsilon (\textbf{k})$ as 
 \begin{equation}
V(r,T)=\int\frac{d^3\textbf{k}}{(2\pi)^{3/2}}
({e^{i\textbf{k}.\textbf{r}}-1})\frac{V(\textbf{k})}{\epsilon(\textbf{k})},
\label{pot_defn}
\end{equation}
where we have subtracted a $r$-independent term (to renormalize the heavy
quark free energy) which is the perturbative free energy of quarkonium at
infinite separation. The dielectric  function is related 
to the ``00''-component of effective gluon propagator in static limit as

\begin{equation}
\frac{1}{\epsilon (\textbf{k})}=\displaystyle
{\lim_{k_0 = 0}}{\textbf{k}}^{2}D_{11}^{00}(k_{0},\textbf{k}),
\label{ek}
\end{equation}

and $V(\textbf{k})$  is the Fourier transform (FT) of the Cornell potential.
To obtain the FT of the potential, we regulate both terms with the same
screening scale. 
However, different scales for the Coulomb and linear pieces were also employed
in~\cite{Megias:IJP85_2011} to include non-perturbative
effects in the free energy beyond the deconfinement temperature through
a dimension-two gluon condensate.

At present, we regulate both terms by multiplying with an exponential
damping factor and is switched off after the FT is evaluated.
This has been implemented by assuming $r$- as distribution
($r \rightarrow$ $r \exp(-\gamma r))$. The FT of
the linear part - $\sigma r\exp{(-\gamma r)}$ is
\begin{eqnarray}
\label{eq-6-3}
-\frac{i}{\textbf{k}\sqrt{2\pi}}\left( \frac{2}{(\gamma-i \textbf{k})^3}
-\frac{2}{(\gamma+i\textbf{k})^3}
\right).
\end{eqnarray}
After putting $\gamma=0$,  we obtain the FT of the linear term $\sigma r$
as,
\begin{equation}
\label{eq-6-4}
\tilde{(\sigma r)}=-\frac{4\sigma}{\textbf{k}^4\sqrt{2\pi}}.
\end{equation}
The FT of the Coulomb piece is straightforward and is given by
\begin{equation}
V_C(\textbf{k})=-\sqrt{(2/\pi)} \frac{\alpha_s}{\textbf{k}^2},
 \end{equation}
thus the FT of the full Cornell potential becomes
\begin{equation}
{V}(\textbf{k})=-\sqrt{(2/\pi)} \frac{\alpha_s}{\textbf{k}^2}-\frac{4\sigma}
{\sqrt{2 \pi} \textbf{k}^4}.
\label{ft_pot}
\end{equation}

The ``00''-component of effective gluon propagator in static limit has been
obtained with the help of ``00''-component of one-loop gluon self energy. 
We have already calculated the ``00'' component of one-loop gluon self energy 
in presence of strong magnetic field at finite temperature in eq.(\ref{g}),
hence the ``00''-component of effective gluon propagator in static limit is 
given by

\begin{equation}
D_{11}^{00}(0,\textbf{k})=\frac{1}{\textbf{k}^2+m_{D}^2}.
\end{equation}

Therefore the real part of the static potential can be obtained
by substituting the dielectric permittivity
$\epsilon(\textbf{k})$ from eq.(\ref{ek}) and the Fourier transformation 
from eq.(\ref{ft_pot})
into the definition of the potential (\ref{pot_defn})
\begin{eqnarray}
V(r;T,B) =V_C(r;T,B)+ V_S(r;T,B),
\end{eqnarray}
where the Coulombic and string term of the potential are given by
(with the dimensionless quantity $\hat{r}=rm_{D}$) 
\begin{eqnarray}
V_{C}(r;T,B)&=&-\alpha_s m_{D}\Big(\frac{e^{-\hat{r}}}
{\hat{r}}+1\Big),\\
V_{S}(r;T,B)&=&\frac{2\sigma}{m_{D}}\Big
[\frac{(e^{-\hat{r}}-1)}{\hat{r}}+1\Big],
\end{eqnarray}
respectively. It is thus evident that the medium dependence in the potential 
enters through the Debye mass, which in turn depends on both temperature
and magnetic field for physical quark masses and depends only on 
magnetic field for massless quarks. This gives a characteristic 
dependence of the potential on both temperature and magnetic field.
The $r$-independent terms in the potential insures $V(r,T)$ to 
reduce to the Cornell potential in $T \rightarrow 0 $ limit~\cite{Agotiya:PRC80'2009}. 
However, such terms could also arise naturally from the basic computations 
of real time static potential in hot QCD \cite{Mikko:JHEP03_2007}
and from the real and imaginary time correlators
in a thermal QCD medium \cite{Beraudo:NPA806_2008}. 
These terms in the potential are needed in computing 
the masses of the
quarkonium states and to compare the results with
the lattice studies. It is equally important while
comparing our effective potential with the free 
energy in lattice studies. 

\begin{figure}[t]
\vspace{-0.4cm}
\begin{center}
\begin{tabular}{c c}
\includegraphics[width=5.5cm,height=5cm]{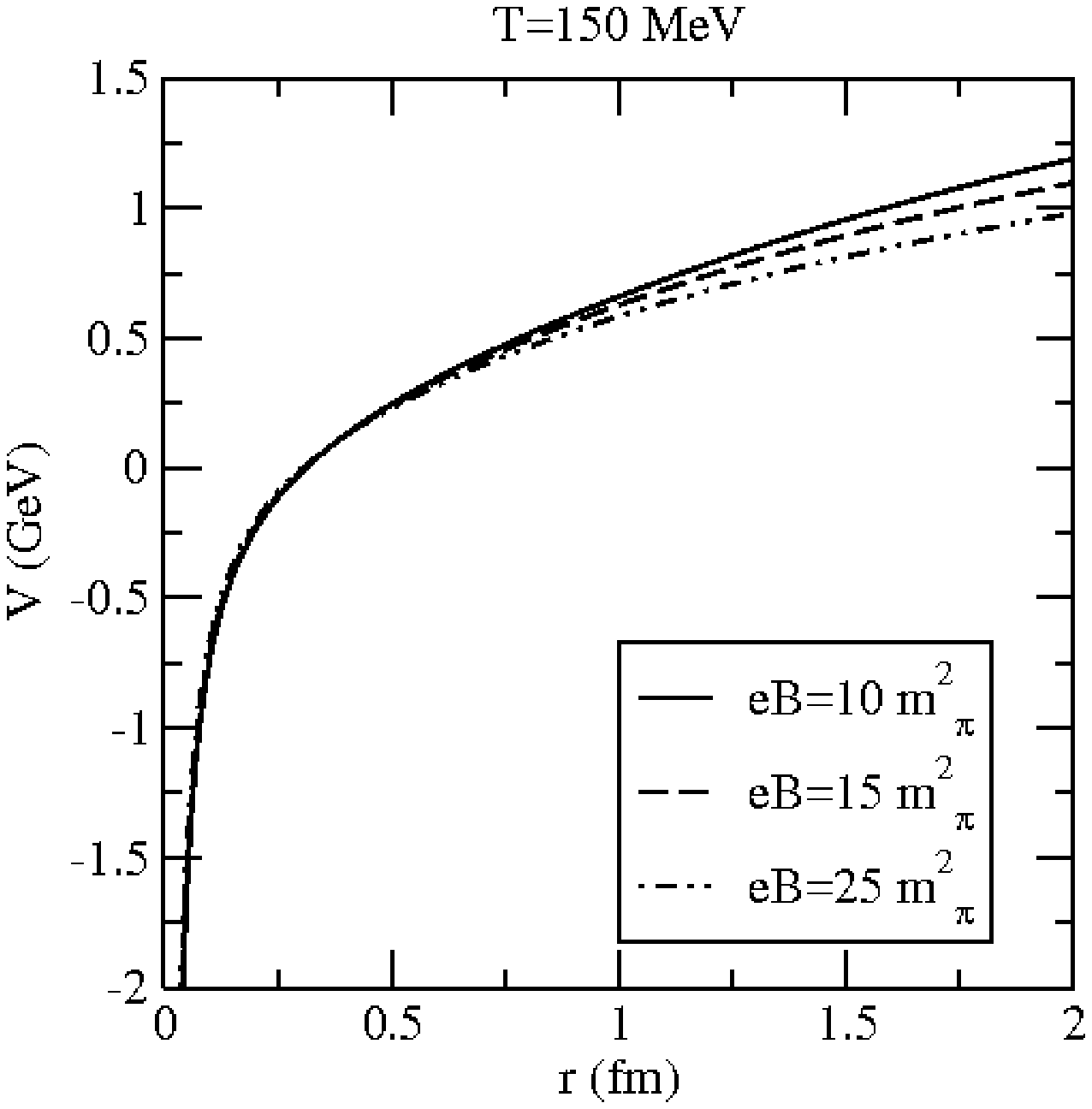}&
\includegraphics[width=5.5cm,height=5cm]{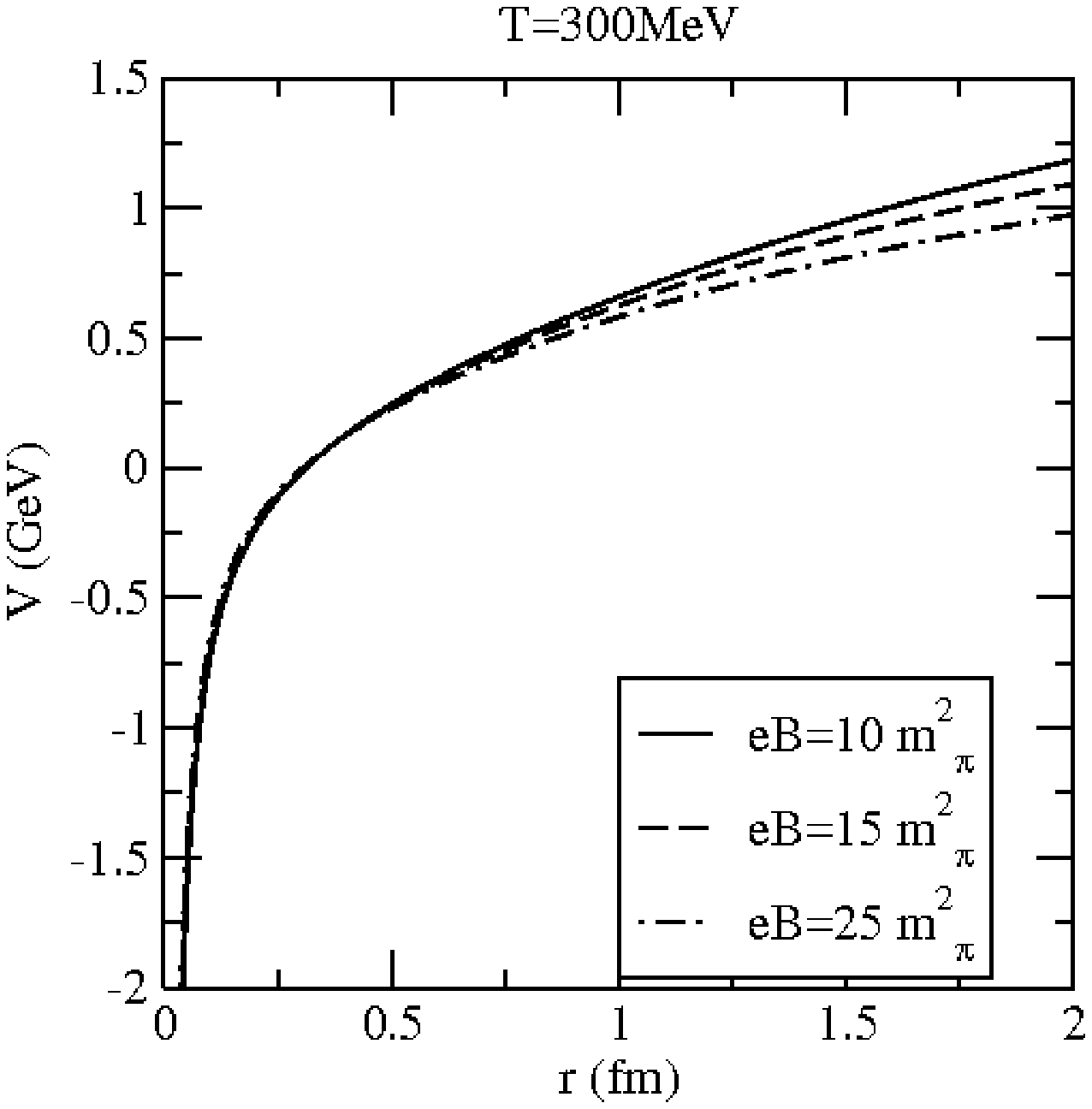}\\
a & b
\end{tabular}
\caption{Effect of magnetic field on potential.}
\end{center}
\label{fig3}
\end{figure}

Since we are exploring the effect of medium on the
potential between $Q$ and $\bar Q$ in strong magnetic field approximation 
so we probe it by varying the strength of magnetic field
($eB$) from $10~m_\pi^2$ to $25~m_\pi^2$ (in Figure 3-a) at a temperature
T=150 MeV. It is found that as the strength of the magnetic field increases 
the potential becomes stronger. To see the competition 
between the magnetic field and temperature, we have plotted the potential 
in Figure 3-b in a hotter medium (T=300 MeV). As we have seen earlier in 
Figure 1-a that the (square) Debye screening mass increases very little 
with temperature, here also the potential changes little as compared to 
Figure 3-a. This small dependence on temperature stems from SMFA as we 
have observed in the Debye screening mass. 

Usually potential model studies are limited to the medium-modification 
of the perturbative part of the potential only where it is assumed
that the string-tension vanishes abruptly at the deconfinement 
point. Since the phase transition in QCD for physical quark masses 
is found to be a crossover~\cite{Karsch_2006},
so the string tension may not vanish at the deconfinement temperature.
This issue, usually overlooked in the
literature where only a screened Coulomb potential was
assumed above $T_c$ and the linear term was neglected,
is certainly worth for an investigation.
To see the effect of the linear term on the potential, in addition to the
Coulomb term, we have plotted the potential (in Fig.4-a) with
($\sigma \neq 0$) and without string term ($\sigma=0$) in a magnetic
field $eB=10~m_\pi^2$. As we know already in vacuum (T=0), the
inclusion of the linear term makes the potential in short-distance
interaction less attractive and in long-distance interaction
the linear term makes the potential more repulsive, compared to
the Coulomb term alone. However, the medium modification causes the linear
term attractive and overall the medium modifications
to both Coulomb and string term makes the potential
more attractive (seen in Figure 4-a) compared to the vacuum potential. 

To see the effect of the scale (Debye mass) at which the screening 
takes place on both the linear and Coulombic term we have plotted 
the potential at a larger magnetic field, $eB=25~m_\pi^2$ 
in Fig. 4-b, where we found that the increase of the scale (screening mass) 
makes the linear term less attractive, compared to the lower scale ($eB$).
To understand the observations in Figure 4, we have probed the range
of interactions, {\em viz.} short-range ($r=0.2$ fm), intermediate
($r=0.5$ fm) and long-range ($r=1$ fm) interactions of $Q \bar Q$
potential as a function of magnetic field ($eB$) in figure 5 
and found that only the long range interaction ($r=1$ fm) has been affected 
noticeably.
Overall observation is that as the strength of the magnetic field 
increases the long range QCD force
becomes more and more short range, thus implying that magnetic field
facilitates early dissolution of $Q \bar Q$ states.

\begin{figure}[t]
\vspace{-0.4cm}
\begin{center}
\begin{tabular}{c c}
\includegraphics[width=5.5cm,height=5cm]{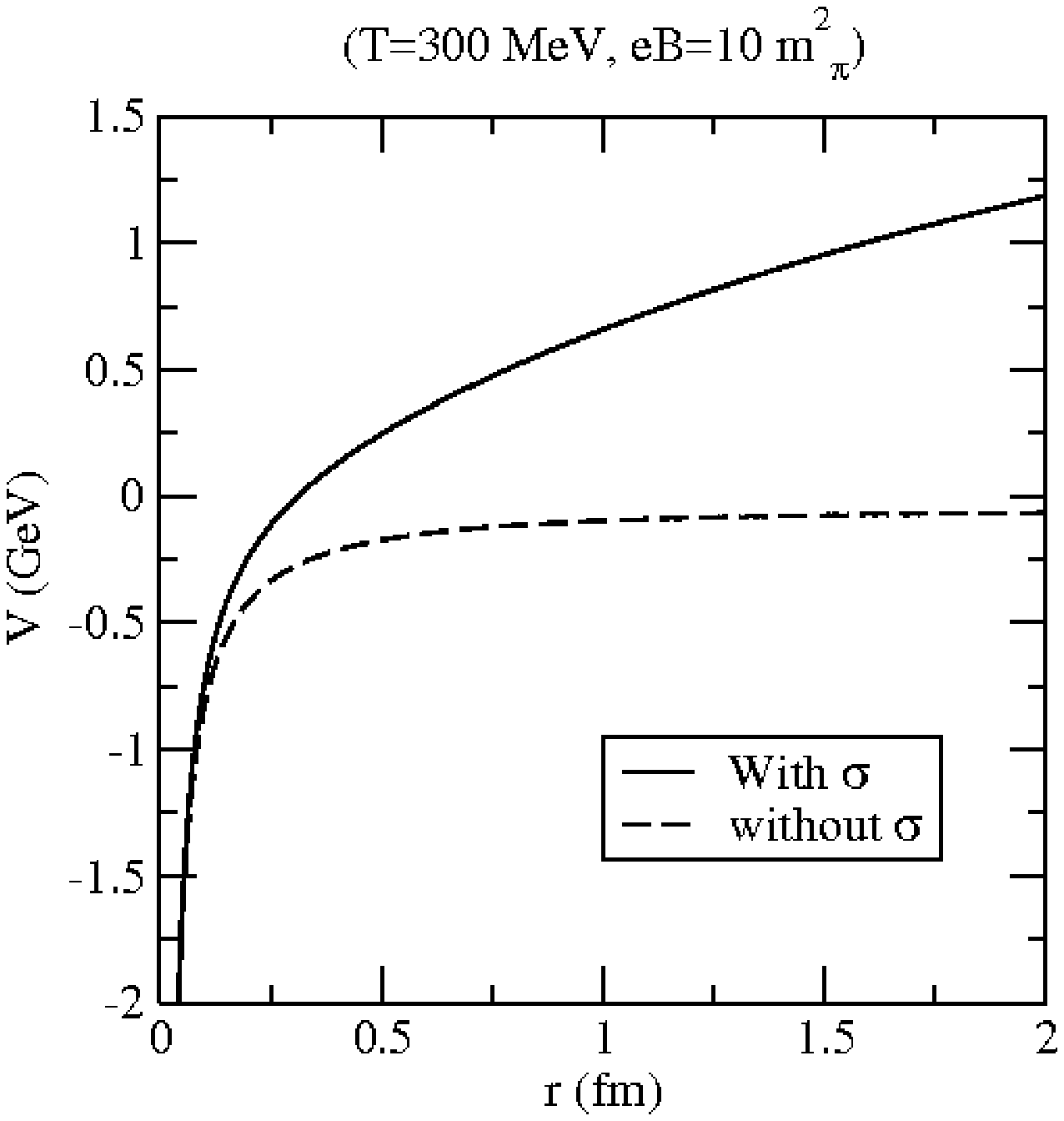}&
\includegraphics[width=5.5cm,height=5cm]{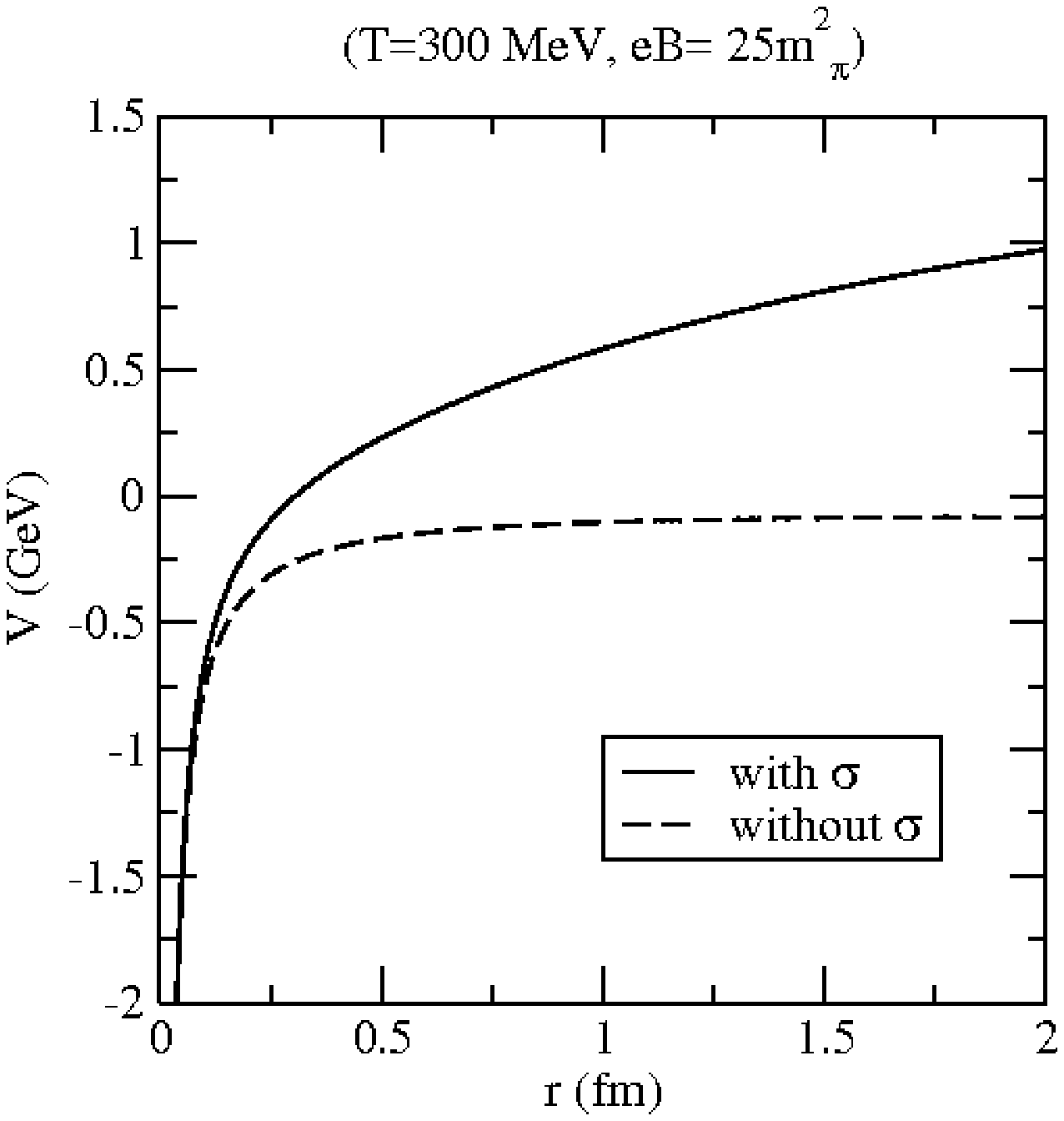}\\
a & b
\end{tabular}
\caption{The effect of string term on potential is depending on magnetic field.}
\end{center}
\end{figure}

\begin{figure}[h]
\vspace{-0.4cm}
\begin{center}
\includegraphics[width=7cm,height=5.9cm]{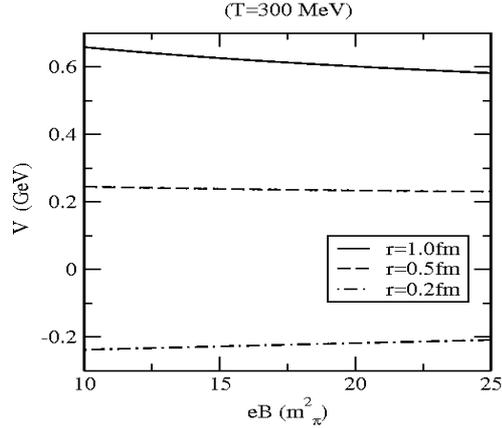}
\caption{Effect of magnetic field on short and long range behavior of the potential.}
\end{center}
\end{figure}
It is important to mention here that we have not observed any anisotropy
in our potential with respect the direction of magnetic field, 
which is expected as a common sense because the magnetic 
field breaks the translational invariance of space. In our perturbative 
framework too, at the starting point the quark propagator
in strong magnetic field approximation gets factorized
into functions involving the parallel and perpendicular components 
with respect to the direction of magnetic field ($\vec{B}=B\hat{z}$) and 
so obviously the gluon self-energy is decoupled into parallel
and perpendicular components. But while thermalizing the quark
propagator and gluon self-energy at finite temperature in strong
magnetic field through the distribution function, there is no scope 
to introduce the momentum anisotropy in distribution function
because the quarks dispersion relation is restricted to the LLL 
only ($E_0(p_z)=\sqrt{p_z^2+m_f^2}$), i.e.
only the longitudinal component of momentum is present, hence no 
anisotropy arises between transverse and longitudinal components. 
One of us had recently derived an anisotropic heavy quark 
potential~\cite{Thakur:PRD88_2013,Thakur:PRD89_2014} in perturbative thermal QCD, where the 
anisotropy in the potential in the coordinate space had arisen from 
the manifested momentum anisotropy with respect to the direction of 
anisotropy in the distribution function. 

However, recently a novel magnetic field-induced anisotropic behavior was 
first observed in the heavy quark potential in the longitudinal-traverse
plane with respect to the direction of $\vec{B}$ \cite{Bonati:PRD89_2014}. 
Later the the set-up is extended to measure Polyakov loop correlators on 
the lattice to extract the potential for both zero and finite temperature 
in place of earlier Wilson loop expectation values used at zero temperature, 
for different orientations with respect to $\vec{B}$ \cite{Bonati:PRD94_2016}. 
The reason of anisotropy arises due to the averaging of both the Wilson 
loop expectation value and Polyakov loop correlators differently for 
different orientations with respect to the magnetic field.

\subsection{Dissociation of Heavy Quarkonia in magnetic field}
In this section, we shall discuss the dissociation of 
charmonium and bottomonium states due to an external strong magnetic 
field in a hot QCD medium. The concept of dissociation temperature becomes 
irrelevant here 
because the scale at which the collective oscillations develop
depends only on the magnetic field albeit we are considering
a hot QCD medium because in strong magnetic 
field approximation ($eB \gg T^2$), the scale at which the collective
oscillation sets in is associated 
with magnetic field only because $eB$ is the most dominant scale
in strong magnetic field limit (if the partons are assumed massless), not the 
thermally generated scales. This in turn makes the potential
to depend only on the magnetic field through the dependence of Debye mass
on the magnetic field. 
Thus, it makes sense here to discuss the dissociation of quarkonium states 
due to the magnetic field only as far as SMFA is valid. 

As we know that in the presence of medium, 
the potential between a heavy quark ($Q$) and its anti-quark ($\bar Q$)
will be screened, as a result if the screening is strong enough,
the potential becomes too weak to form the resonance. Thus we can argue 
that the quarkonium states will be dissolved in a medium if 
the Debye screening radius, $r_D$ (=$\frac{1}{m_D}$) in a given medium
is smaller than the bound state radius of a particular resonance state
then the medium inhibits the formation of the particular resonance and
$Q$ and $\bar Q$ will be dissolved into the medium.
Since the screening mass in strong magnetic field increases with the 
magnetic field therefore the (critical) magnetic field at which
the $Q \bar Q$ potential becomes too feeble to hold $Q \bar Q$ together
becomes smaller for the excited states. We can thus estimate
the lower limit of critical magnetic field for various charmonium and 
bottomonium states by the criteria: $\sqrt{\langle {r^i}^2 \rangle}= 
{r_D} (B_d^i)$, i.e., for magnetic field larger than ${B_d}^i$, 
the $i$-th quarkonium states cease to exist. 
For example, $J/\psi$ will be dissociated at $eB=14 m_\pi^2$ and its 
excited state, $\psi^\prime$ is dissociated at smaller magnetic field 
$eB= m_\pi^2$ whereas $\Upsilon$ will be dissociated at $eB=130 m_\pi^2$ and 
$\Upsilon^\prime$ is dissociated at smaller magnetic field $eB= 13 m_\pi^2$.

To understand the in-medium properties of the quarkonium 
states quantitatively, one need to solve the Schr\"{o}dinger equation with the 
medium-modified potential, $V(r;B,T)$. There are some numerical 
methods to solve the Schr\"odinger equation
either in partial differential form (time-dependent) or eigen value form
(time-independent) by the
finite difference time domain method (FDTD) or matrix method, respectively.
In the later method, the stationary Schr\"odinger equation
can be solved in a matrix form through a discrete basis, instead
of the continuous real-space position basis spanned by the states
$|\overrightarrow{x}\rangle$. Here the confining potential V is subdivided
into N discrete wells with potentials $V_{1},V_{2},...,V_{N+2}$ such that
for $i^{\rm{th}}$ boundary potential, $V=V_{i}$ for $x_{i-1} < x < x_{i};
~i=2, 3,...,(N+1)$. Therefore for the existence of a bound state, there
must be exponentially decaying wave function
in the region $x > x_{N+1}$ as $x \rightarrow  \infty $ and
has the form:
\begin{equation}
\Psi_{N+2}(x)=P_{{}_E} \exp[-\gamma_{{}_{N+2}}(x-x_{N+1})]+
Q_{{}_E} \exp [\gamma_{{}_{N+2}}(x-x_{N+1})] ,
\end{equation}
where, $P_{{}_E}= \frac{1}{2}(A_{N+2}- B_{N+2})$,
$Q_{{}_E}= \frac{1}{2}(A_{N+2}+ B_{N+2}) $ and,
$ \gamma_{{}_{N+2}} = \sqrt{2 \mu(V_{N+2}-E)}$. The eigenvalues
can be obtained by identifying the zeros of $Q_{E} $. Using this 
method, we have found that $J/\psi$ and $\Upsilon$ is dissociated 
at $eB=5 m_\pi^2$ and $eB=50 m_\pi^2$, respectively.

Though the dissociation magnetic fields, obtained from the 
two different methods apparently look different, its easy to see 
that qualitatively they are similar. Using both methods, we found 
that the dissociation magnetic field for $\Upsilon$ is roughly an 
order of magnitude greater than the dissociation magnetic field 
for $J/\psi$. Even though their absolute value obtained from the 
two different methods differ, they lie in the same ball park, 
which is $\sim$ 10 $m_\pi^2$ for $J/\psi$ and $\sim$ 100 $m_\pi^2$ 
for $\Upsilon$.

\section{Conclusions}

In this article, we have explored the effects of strong and homogeneous 
magnetic field on the properties of quarkonium states. For that purpose
we have derived the potential between a heavy quark and its anti-quark 
by the medium corrections to both Coulomb and linear term of 
$Q\bar Q$ potential at T=0, {\em unlike the medium correction to the Coulomb 
term alone}. Although the medium considered is thermal but 
due to strong magnetic field approximation, all other scales present in the 
thermal medium becomes irrelevant as the scale related to
magnetic field dominates over other. This
is exactly what happens in the collective oscillation of the
medium in the form of Debye mass. In fact, the Debye mass
becomes completely independent of temperature for massless 
quarks and depends very weakly on temperature for massive quarks. 
However, beyond a certain temperature, the dependence is so weak 
that it is almost insignificant.
As a result the heavy quark potential mainly depends on
the magnetic field with a very feeble dependence on the
temperature. This is expected as the effect of the medium on
the potential enters through the Debye mass. In particular 
the long distance part of the potential gets significantly affected, 
whereas the short distance part is mildly affected.

We have then studied the dissociation of quarkonium states in a medium. 
Since the potential in SMFA depends mainly only on the magnetic field 
thus we have discussed the dissociation of quarkonium states
due to the magnetic field only. We have estimated 
the critical value of magnetic field beyond which the 
resonance does not form in two methods.
The first one gives a lower limit of critical magnetic field for 
both charmonium and bottomonium states at which 
the Debye screening radius becomes smaller than the bound state 
radius of a particular resonance state. The other one
comes from the consideration of the binding energies of a specific 
state obtained from the energy eigenvalues of the Schr\"odinger equation.
In brief, $J/\psi$ is dissociated at $eB \sim$ 10 $m_\pi^2$ and 
$\Upsilon$ is dissociated at $eB \sim$ 100 $m_\pi^2$. 

\section{Acknowledgments}
We are thankful to Aritra Bandyopadhyay for a fruitful discussion 
during this work. Bhaswar is thankful to the Ministry of 
Human Resource Development, 
Government of India for the financial assistance.

\end{document}